\documentclass[useAMS,usenatbib]{mn2e}
\usepackage{graphicx}

\title[Inclination impact for imaging of Exo-Earths]{The relevance of prior inclination determination for direct imaging of Earth-like planets}
\author[M. Janson]
{
M. Janson$^{1}$\thanks{Reinhardt fellow. E-mail: janson@astro.utoronto.ca}
\\
\\
$^{1}$Department of Astronomy, University of Toronto, 50 St George St, Toronto, M5S 3H4, Canada}
\begin{document}

\date{N/A}

\pagerange{\pageref{firstpage}--\pageref{lastpage}} \pubyear{2010}

\maketitle

\label{firstpage}

\begin{abstract}
Direct imaging and characterization of extrasolar Earth-like planets is strongly impacted by the orbital inclination of the planet to be studied, as a combination of pure geometrical effects and the impact of exozodiacal dust. Here, we perform simulations to quantify the impact of a priori knowledge of inclination for the efficiency of a typical coronagraphic or occulter-based mission. The relative impact and complementarity with prior knowledge of exozodiacal brightness down to achievable levels is examined and discussed. It is found that inclination has an even greater impact than the exozodiacal brightness, though the two have excellent complementarity. We also discuss different methods for inclination determination, and their respective applicability to the context of precursor science to an imaging mission. It is found that if technologically achievable, a combined effort to determine inclinations and exozodiacal brightnesses with ground-based facilities would substantially increase the efficiency of a space-based dedicated mission to image and characterize Earth-like planets.
\end{abstract}


\section{Introduction}
\label{sec_intro}

The planar geometries of planetary systems are of fundamental importance for the detection and characterization of planets. Radial velocity is biased towards systems that are close to edge-on, and the typically unknown inclination leads to an uncertainty in mass. Transiting systems can only be observed under very tight inclination boundaries. In direct imaging, a planet on a highly inclined orbit will spend at least part of its time inside of the inner working angle (IWA) of the working instrument. Gravitational microlensing detections favour projected separations close to the Einstein ring radius, hence projection effects from inclination will play a role also in this case. By contrast, the detection efficiency of astrometry can be said to be relatively independent of inclination (e.g. Unwin et al. 2008).

It follows from this importance of orbital inclination for detection with a wide range of methods, that an a priori determination of this quantity could potentially lead to a significant boost in detection efficiencies of (otherwise) blind surveys. An indirect way of doing so would be to determine the rotational plane of the star. Since a star and its disk form primordially from the gravitational collapse of a rotating cloud with conservation of angular momentum, the plane of the disk should in general be similar to the rotational plane of the star, and so the difference in the final inclination of planets formed in the disk and the stellar rotation should be small. Indeed, we know that this is the case in our own Solar system (e.g. Beck \& Giles 2005). We also know from systems like $\epsilon$ Eri (Saar \& Osten 1997; Greaves et al. 2005) and Fomalhaut (Le Bouquin et al. 2009) that the rotational plane of the star is consistent with the disk plane of the circumstellar material. A projection of the mutual inclination can also be measured directly in transiting systems through the Rossiter-McLaughlin effect. The general population of studied hot Jupiters so far is consistent with coplanarity (e.g. Fabrycky \& Winn 2009). However, some systems do exist in which the planet is markedly non-coplanar (e.g. Narita et al. 2009, Winn et al. 2009). Probably, a violent dynamical history or Kozai influence from a stellar companion is responsible in these cases.

A study by Beatty \& Seager (2010) investigated the impact of a priori known inclinations for the study of transits among nearby stars. Partly motivated by this study, we here investigate the impact of known inclinations for the purpose of direct imaging of Earth-like planets. Direct imaging strongly benefits from low inclinations (as opposed to transits, where high inclinations are necessary). The reasons for this are twofold: Consider an edge-on system ($i = 90^{\rm o}$) and a face-on system ($i = 0^{\rm o}$), at the same distance and with an equal planet at equal semi-major axis on a circular orbit. The semi-major axis is such that, at maximum projected separation, the planet is outside of the observer's IWA. In the face-on case, the planet is then permanently visible, but in the edge-on case, for some fraction of the time the planet is inside of the IWA and will remain undetected if observed at an unfortunate epoch. Furthermore, however, the difference in detectability is emphasized to an even greater extent if the detection limit is set by exozodiacal dust. In the edge-on case, a larger portion of the exozodiacal emission is projected into the line-of-sight such that at maximal projected separation, the detection limit is significantly worse than for the face-on case. The edge-on planet spends the majority of its time well within the maximal projected separation, where the exozodiacal emission is even yet worse.

In this paper, we quantify these effects for typical observational parameters of dedicated Earth-like exoplanet imaging missions in the visual range. The conclusions apply equally to occulter-based (e.g. Cash 2006; Janson 2007) and coronagraph-based (Guyon et al. 2006) designs, although some practical differences exist which we will point out as they occur. The inclination is also of importance for infrared interferometric designs (e.g. Cockell et al. 2009), but these will not be considered in this paper. In the following, we will describe the input parameters to the simulations and how the simulations were performed, followed by the results and discussion. Some viable and non-viable methods for determining inclinations are discussed before the conclusion.

\section{Efficiency simulation}
\label{sec_simulations}

We have performed simulations to test the impact of knowing the inclination distributions a priori, relative to lacking this knowledge. The idea is to simulate the relative time it would take to detect an Earth-like planet against an exozodiacal background at different points of time, for various inclinations, exozodi brightnesses, and target distances, as well as the fractional time that the planet is within and outside of the IWA, respectively. These quantities are combined to quantify a relative observation efficiency between different targets under different circumstances. Different target selections are then made, which are blind to the inclinations or to various degrees of exozodiacal emission, and are compared to each other and to the perfect-knowledge case to determine the impact of various degrees of knowledge on the total efficiency of the survey.

\subsection{Input parameters and distributions}
\label{sec_input}

The only constraints put on the observing facility are an IWA of 60 mas outside of which a contrast of $10^{-10}$ -- sufficient for an Earth-like planet to dominate the residual stellar signal -- is reached, and a FWHM of 20 mas, corresponding roughly to a 6-meter telescope at 600 nm. The most important parameter of the target sample is the distance, and so for a distance distribution of our simulated target stars we choose all targets from HIPPARCOS (Perryman et al. 1997) within 16.7 pc with $0.4 < B-V < 1.0$ to represent reasonably solar-like conditions. For simplicity, the stars are all assumed to have the properties of a 1$M_{sun}$ star, although in reality, the colour selection spans 0.7-1.3$M_{sun}$. Here the outer distance of 16.7 pc is simply set by the fact that beyond this distance, the angular separation corresponding to 1 AU is smaller than the IWA. Every star is assumed to have exactly one Earth-like planet at a circular 1 AU orbit with a star-planet contrast of $3.34*10^{-10}$ (e.g. Janson 2007). The random distribution of inclinations is set as the arccosine of a uniform random distribution, to accurately describe the real distribution. The number of stars fulfilling the distance and colour selection criteria is 164, and we will assume that 60 of those would be selected for the mission. The optimal number of stars in an actual survey depends on several parameters, particularly including the fraction of systems that host habitable planets, $\eta _{\earth}$. If $\eta _{\earth}$ is known a priori, it is possible to optimize the number of targets for a given mission concept (e.g. Beichman 2003).

For the statistical brightness distribution of exozodiacal dust disks, we use a modified single-population equation from Greaves \& Wyatt (2010):

\begin{equation}
Z(x) = cx^{-{\alpha}}
\end{equation}

where $x$ is the rank of the star such that among 100 stars, the brightest disk has $x = 0.01$ and the faintest has $x=1.00$, and $c$ is a normalization constant such that $Z(0.9) = 1$ zodi. The latter normalization means that at the rank of the Sun, which is assumed to have rank 0.9, the brightness is 1 zodi. 

\begin{figure}
   \centering
   \includegraphics[width=8.0cm]{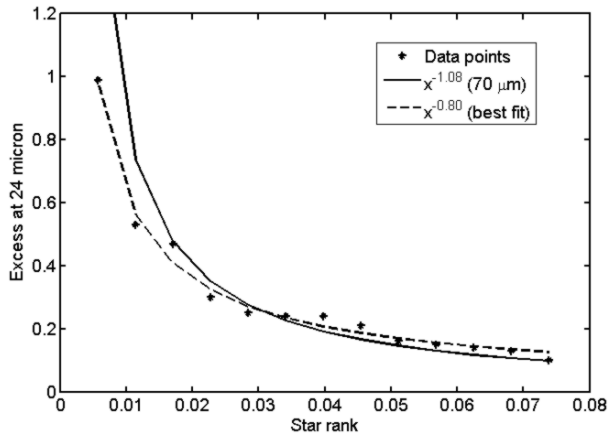}
\caption{Infrared excess emission at 24 $\mu$m for known targets (stars) in the literature. The solid line corresponds to the $x^{-1.08}$ power law derived from 70 $\mu$m excesses. A best-fit to the full ensemble (dashed line) is provided by $x^{-0.80}$ (see the text for discussion).}
\label{slope24}
    \end{figure}

The original exponent in Greaves \& Wyatt (2010) was $\alpha = 1.08$. However, this relation was derived from infrared excess at 70 $\mu$m, which probes cold dust in the outer disk, and hence a different region of the disk from what we are concerned with here. A better indicator for warm dust in the inner disk is excess at 24 $\mu$m, and hence this wavelength range is preferable for our purpose of estimating the brightness distribution of exozodi. Excess at 24 $\mu$m is often harder to detect than at 70 $\mu$m because of the more demanding contrast to the stellar photosphere, but some detections were made by Beichman et al. (2006) and Trilling et al. (2008), the samples that also form the basis for the 70 $\mu$m excesses used by Greaves \& Wyatt (2010). In total, 13 systems exist which have 24 $\mu$m excess with a confidence of $>$3$\sigma$ in these samples. Recently, Koerner et al. (2010) presented 29 systems with 24 $\mu$m excess, but these are primarily around K-type stars, whereas we are concerned with Sun-like stars. The Koerner et al. (2010) sample lacks systems of the order of the highest excesses in Beichman et al. (2006) and Trilling et al. (2008), which should be those that are easiest to detect. Hence, we consider that there is a risk that the low-mass sample will bias the result and do not use this sample for the main simulation. We plot the 13 systems in Fig. \ref{slope24}, along with the $\alpha = 1.08$ power law. The fit is reasonable, but cannot match the two highest excess cases. The full ensemble of stars can be fit better by using a somewhat shallower slope, the best fit is provided by $\alpha = 0.80$, also plotted in the figure. Hence, we use this exponent in the simulations, although it is based on fewer targets that the relation fit in Greaves \& Wyatt (2010), since 24 $\mu$m excess is more reasonable for the exozodiacal dust we are concerned with. 

It should be emphasized that even the original distribution is an extrapolation based on observations in a very limited parameter range. Hence, the distribution is very uncertain, but represents a "best guess" case on the basis of the existing observations. Note that alternative simulations based both on the 70 $\mu$m excesses and the Koerner et al. (2010) 24 $\mu$m excesses are presented in Sect. \ref{sec_alternative}.

The face-on surface density distribution of each exozodiacal disk is another uncertainty, but here it is assumed to be $r^{-0.3}$ out to 3 AU. Studies in the visual wavelength range have implied that this is the case for the zodiacal disk at 0.3-1 AU, assuming a thin wedge-like structure (Leinert et al. 1981), and this holds approximately true also at $\sim$1-3 AU (Hanner et al. 1976). The reflected emission goes as $r^{-2.0}$, so in total the brightness distribution goes as $r^{-2.3}$. However, a recent study at 24 $\mu$m has given a brightness distribution of about $r^{-3}$ (Nesvorn\'y et al. 2010). Assuming that the dust temperatures are much higher than 120 K such that the radiation is in the Rayleigh-Jeans regime, the temperature of the dust emission will scale as $r^{-0.5}$, which leads to a density distribution of $r^{-2.5}$, which in turn leads to a visual brightness distribution of $r^{-4.5}$, vastly different from what we have assumed above. Since we are primarily interested in visual rather than infrared radiation here, we adopt the former value for the main simulation. However, to emphasize the effect of this large discrepancy, we also include a simulation based on the latter assumption in Sect. \ref{sec_alternative}.

For normalization, the surface brightness of the zodiacal cloud at 1AU (1 zodi) is 23 mag (arcsec)$^{-2}$ (e.g. Absil et al. 2009). The disk is assumed to be azimuthally flat. In reality, the dynamical influence of planets in the disk will introduce asymmetric features. A simulation based on the Earth case has been performed by Dermott et al. (1994). In those simulations, it was found that the excess flux from the strongest morphological feature (a density enhancement in an Earth-trailing orbit) is only 10\%. Since our simulations span about a factor 1000 in disk brightnesses, this effect is negligible in comparison, and can be disregarded. The effect that it would have, if stronger, would be maximal for a disk seen edge-on, and minimal for face-on.

The contribution of zodiacal dust from the Solar system itself within the telescope beam ($f_{\rm z}$) is assumed to be 25\% of the brightness of a 1 zodi edge-on exozodiacal disk, following the values used in Agol (2007).

\subsection{Procedure}
\label{sec_proc}

For 1000 populations $p$ and for 100 instances of time $t$ per population, we generate random inclinations $i$ and total exozodiacal brightnesses $z$ for the 164 targets according to the distributions described above. At each instance, the planet is located at a projected separation 

\begin{equation}
\rho (t) = \sqrt{(\cos \phi (t))^2 + (\sin \phi (t) \cos i)^2}
\end{equation}

where $\phi (t)$ is the phase angle at some instance of the orbit. The projected separation is either outside of or inside of the IWA, setting $f_{{\rm vis},t}$ to one or zero, respectively. The average over instances $\bar{f}_{\rm vis}$ is the fractional time of an orbit during which the planet is visible.

For each instance when the planet is visible, the line-of-sight brightness of the exozodiacal dust $f_{{\rm ez},t}$ must be calculated at the position of the planet. This is done by first calculating the surface brightness for a face-on disk as a function of physical position in two dimensions, in steps of 0.002 AU out to 3 AU. The flux is then evaluated in a box corresponding to a 20 by 20 mas area. The box is centered on the x-axis, at a separation given by the projected separation of the planet at a given instance. A vertical stretch is also applied to the box to account for the tilt of the disk relative to the observer. In other words, scaling as function of distance and projection as function of inclination are done by the corresponding inverse scaling and de-projection of the box. The flux is normalized by the randomly assigned $z$ factor, and is finally normalized by $d^{-2}$.

   \begin{figure}
   \centering
   \includegraphics[width=8.0cm]{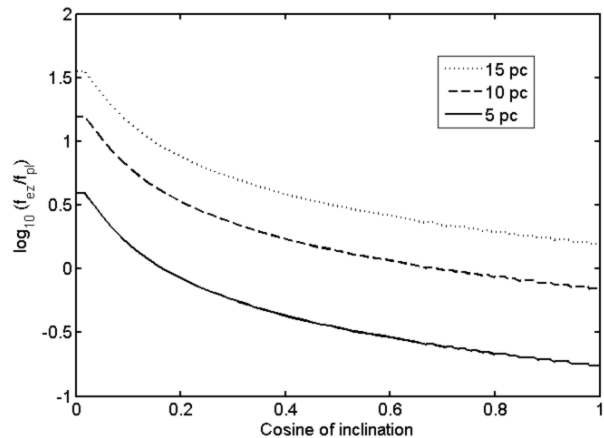}
\caption{Flux fraction between an Earth-like planet and an exozodiacal disk as function of inclination, at distances of 5 pc (solid line), 10 pc (dashed line), and 15 pc (dotted line), at a projected separation of 1 AU. Even though the fraction is a strong function of distance, it is still the case that a very small inclination system at 15 pc gives a better sensitivity than a very high inclination 5 pc system, even at the maximum projected separation (1 AU). The inclination is plotted in a $\cos i$ scale to represent uniform probabilities. The edge-on case is $\cos i = 0$ and the face-on case $\cos i = 1$. The flattening close to edge-on is due to the 3 AU outer cut-off of the exozodiacal disk.}
\label{zodidist}
    \end{figure}

The planetary brightness $f_{{\rm pl},t}$ is calculated from the $3.34*10^{-10}$ contrast to an absolute magnitude of a Sun-like star of $M_{\rm V}=4.7$ mag, normalized to the distance of the system. Furthermore, we factor in an illumination function modeled by $I(t) = 1 + \sin(\phi (t)) \sin(i)$ to account for the different phases of the planet during an orbit. Given both the exozodiacal and planet brightness, we can calculate a signal-to-noise ratio as 

\begin{equation}
\left(\frac{S}{N}\right)_t = \frac{f_{{\rm pl},t}}{\sqrt{f_{{\rm pl},t} + f_{{\rm ez},t} + f_{\rm z} + f_{\rm psf}}} .
\end{equation}

Here, $f_{\rm psf}$ is the PSF brightness outside the IWA, which is $10^{-10}$ of the stellar flux, as discussed above. From this we can calculate a relative time $\tau_t = (S/N)_t^{-2}$ required to reach some given sensitivity. With this parameter in hand, we can calculate a total inverse observing efficiency as $\xi = \tilde{\tau}/\bar{f}_{\rm vis}$ for each target in each population. To summarize, the parameter $\xi$ corresponds to how long (relative to other targets) we would need to integrate to achieve an acceptable significance for detection, with the inverse probability that the target will be observable in the first place (i.e., outside of the IWA) factored in. Note that we have chosen the median rather than the mean of $\tau$, this is because in nearby high-inclination systems, the rare situation can arise that a planet is formally visible (outside of the IWA), but the phase angle is such that hardly any flux reaches the observer, which leads to an absurdly long required integration time. When taking the mean, this rare instance affects all other instances to a dubious extent. Although it is formally correct under the assumption that the observer keeps integrating until finding an Earth-equivalent planet, or otherwise integrates long enough to have adequate sensitivity to detect the faintest Earth-equivalent object in the field, what would more realistically happen in these cases is that the observer would cut losses by ending the observation prematurely. Hence, the median is a more robust estimator for our purposes, although it may somewhat under-estimate the positive impact of prior knowledge.

Using $\xi$, we can select the 60 best targets for a survey. If we had perfect knowledge of the inclination and the exozodiacal brightness, we could make the perfect selection to minimize $\Sigma$, the sum of $\xi$ for the selected 60 targets in every randomly generated population. We denote the average $\Sigma$ resulting from a target selection with perfect knowledge over all populations as $\bar{\Sigma}_{\rm ideal}$. 

   \begin{figure}
   \centering
   \includegraphics[width=8.0cm]{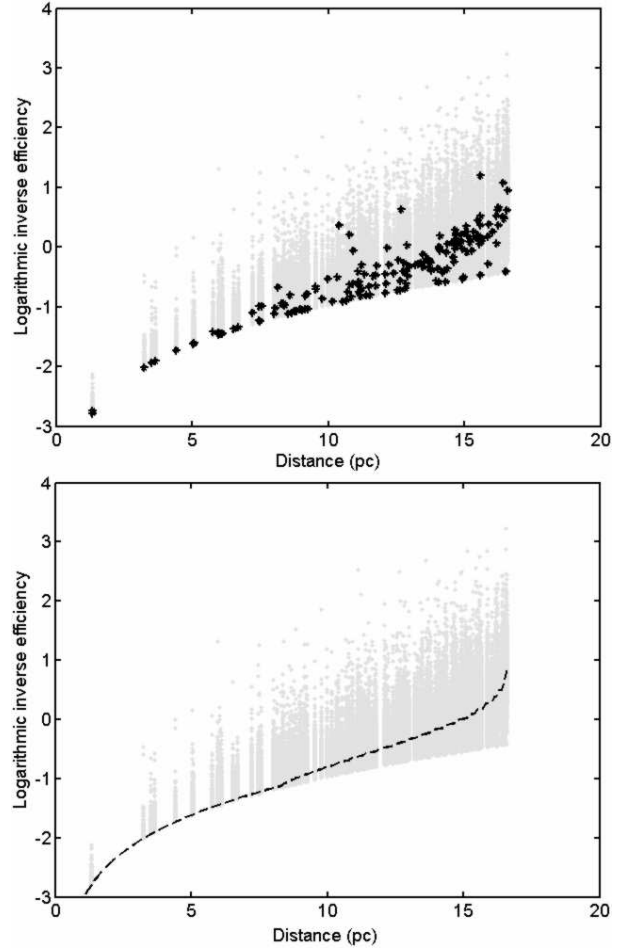}
\caption{Dependence of $\xi$ on distance $d$. The gray dots correspond to all targets in all populations of our simulation. Upper: The black symbols are the targets of one random population. The few systems that exist within $\sim$5 pc are exceptional and will yield a low $\xi$ almost regardless of $z$ and $i$. Outside of this, the other parameters start to strongly influence $\xi$. Lower: The dashed line shows the dependency on $d$ for a hypothetical target with $z=1.6$ zodi and $i=60^{\rm o}$.}
\label{xidist}
    \end{figure}

   \begin{figure}
   \centering
   \includegraphics[width=8.0cm]{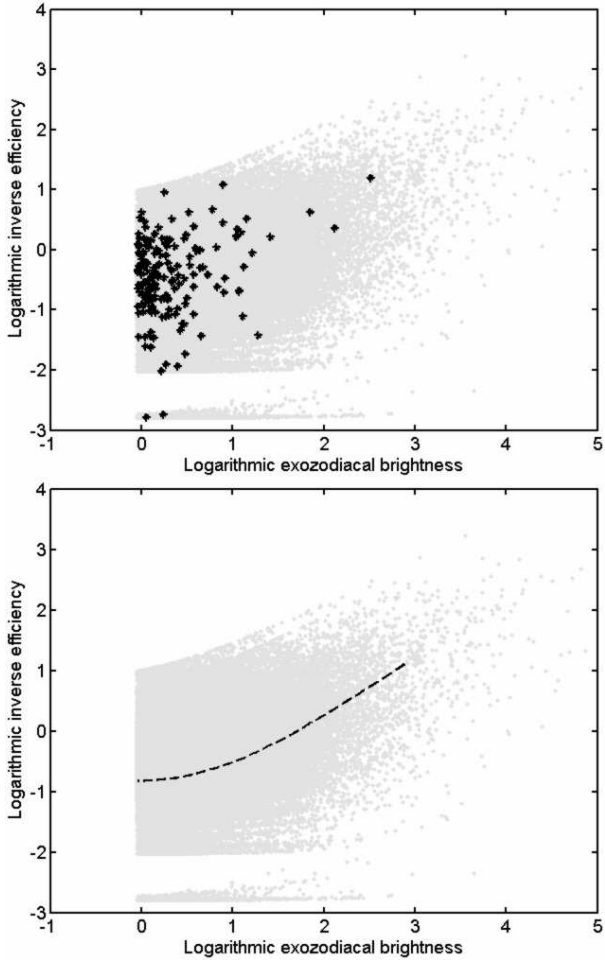}
\caption{Dependence of $\xi$ on exozodiacal brightness $z$. The gray dots correspond to all targets in all populations of our simulation. Upper: The black symbols are the targets of one random population. This parameter is particularly important above $z$ $\sim$10 zodi, below this limit the $\xi$ dependence is relatively weak. Lower: The dashed line shows the dependency on $z$ for a hypothetical target with $d=10$ pc and $i=60^{\rm o}$.}
\label{xizodi}
    \end{figure}

   \begin{figure}
   \centering
   \includegraphics[width=8.0cm]{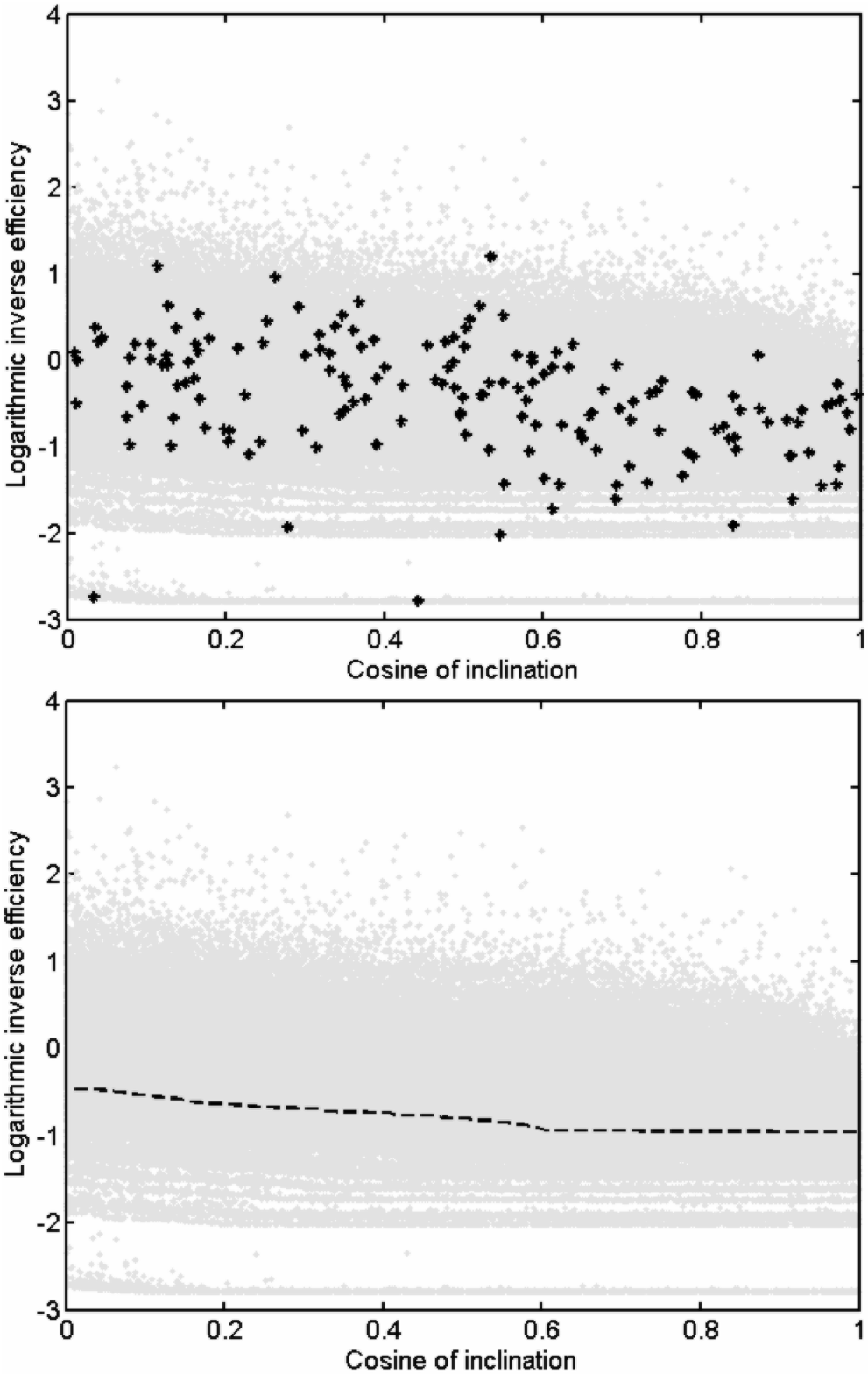}
\caption{Dependence of $\xi$ on inclination $i$. The gray dots correspond to all targets in all populations of our simulation. Upper: The black symbols are the targets of one random population. The inclination is plotted in a $\cos i$ scale to represent uniform probabilities. The edge-on case is $\cos i = 0$ and the face-on case $\cos i = 1$. There is a gradual improvement of $\xi$ over the whole range from edge-on to face-on systems. Lower: The dashed line shows the dependency on $i$ for a hypothetical target with $d=10$ pc and $z=1.6$ zodi.}
\label{xiincl}
    \end{figure}

The relative efficiency resulting from different degrees of actual knowledge can now be evaluated by systematically losing information and calculating the $\bar{\Sigma}$ that results from the corresponding target selections. We assume that there are three degrees of knowledge for $z$. The worst knowledge is if we only know about all systems with $z \geq 10^3$ zodi. This roughly corresponds to our present knowledge, and is denoted as case `z3'. The mid-level case of knowledge is when all systems with $z \geq 10^2$ zodi are known, denoted `z2'. This represents a reasonable goal for a near-term dedicated effort to characterize exozodiacal disks. Case `z1' with all $z \geq 10^1$ zodi systems known is the best case, and would require a highly substantial effort, but could be possible using the Large Binocular Telescope Interferometer (according to Absil et al. 2008). For inclination, we assume that either no inclinations are known (case `i0'), or that all inclinations are known (case `i1'). This is also a reasonable distinction, since except for a few special cases, the observational case for determining the inclination (discussed in Sect. \ref{sec_inclin}) should be about equal for nearby Sun-like stars, such that if the inclination can be determined in one case, it can be determined in all cases. 

A pseudo-merit $\Psi$ corresponding to the real merit $\Sigma$ is calculated for all cases of limited knowledge. This is done by redoing the simulations and replacing every unknown quantity with the expected mean value of the statistical distribution. For instance, in calculating $\Psi_{\rm z2,i0}$, every $i$ is replaced with 60$^{\rm o}$, and every $z < 100$ zodi is replaced with the value $Z(0.5) = 1.6$ zodi. A selection of 60 targets per population is done to minimize $\Psi_{\rm z2,i0}$, and then the $\bar{\Sigma}_{\rm z2,i0}$ of the real distributions is calculated on the basis of that selection. In this way, the $\bar{\Sigma}$ of each case represents the real relative inverse efficiency based on a best-effort selection using the knowledge at hand. As a final step, we define the relative efficiency $\epsilon_{\rm case} = \bar{\Sigma}_{\rm z3,i0} / \bar{\Sigma}_{\rm case}$. The normalization is done so that $\epsilon_{\rm z3,i0} = 1.00$ is the present-day efficiency. 

\subsection{Results}
\label{sec_results}

The values of each $\epsilon_{\rm case}$ are listed in Table \ref{inctab1}. A particularly notable fact is that $\epsilon_{\rm z3,i1} > \epsilon_{\rm z1,i0}$, i.e., knowing the inclinations has a larger impact on the mission efficiency than determining exozodiacal dust brightnesses to an achievable degree. The very best efficiency is achieved by doing both: $\epsilon_{\rm z1,i1}$ is even close to $\epsilon_{\rm ideal}$. The fact that $(\epsilon_{\rm z3,i1} - \epsilon_{\rm z3,i0}) + (\epsilon_{\rm z1,i0} - \epsilon_{\rm z3,i0})$ is close to $(\epsilon_{\rm z1,i1} - \epsilon_{\rm z3,i0})$ signifies that there is considerable complementarity between the two factors, in the sense that one does not replace the other, but a maximal effort in both areas is what gives the optimal performance. Also notable is the fact that $\epsilon_{\rm z3,i1}$ is 23\% higher than $\epsilon_{\rm z3,i0}$. Since a program for determining the inclinations (if performed e.g. in the form of SONG, see Sect. \ref{sec_large}) is very likely to have significantly less than 20\% of the cost of a dedicated mission searching for extrasolar Earth-like planets, it would clearly be a good investment to execute such a program in this context.

\begin{table}
   \centering
\caption[]{Relative efficiencies for various degrees of knowledge.}
         \label{inctab1}
\begin{tabular}{lll}
            \noalign{\smallskip}
            \hline
            \noalign{\smallskip}
$z$ knowledge & $i$ unknown & $i$ known \\
            \noalign{\smallskip}
            \hline
            \noalign{\smallskip}
$z > 1000$ & $\epsilon_{\rm z3,i0} = 1.00$ & $\epsilon_{\rm z3,i1} = 1.23$ \\
$z > 100$  & $\epsilon_{\rm z2,i0} = 1.06$ & $\epsilon_{\rm z2,i1} = 1.28$ \\
$z > 10$   & $\epsilon_{\rm z1,i0} = 1.12$ & $\epsilon_{\rm z1,i1} = 1.32$ \\
Ideal case & N/A                           & $\epsilon_{\rm ideal} = 1.33$ \\
            \noalign{\smallskip}
            \hline
\end{tabular}
\end{table}

The primary dependency on distance is due to the fact that within a fixed spatial sampling, the planetary flux decreases with $d^2$, whereas the disk brightness remains constant. For this reason, the few targets inside of 5 pc are exceptionally good and will almost never decrease the average efficiency, regardless of the disk brightness and inclination. However, at 5 pc, the disk-to-planet contrast is already less favorable in the edge-on case than at 15 pc in the face-on case, even at the maximum projected separation of 1 AU. This is illustrated in Fig. \ref{zodidist}. In the edge-on case, the contrast becomes yet worse as the projected separation decreases. Breakdowns of the dependency of the inverse efficiency $\xi$ on $d$, $z$, and $i$ are illustrated in Figs. \ref{xidist}, \ref{xizodi}, and \ref{xiincl}, respectively. As mentioned, the distance is particularly important at small $d$. The exozodiacal brightness has the largest impact on $\xi$ when $z > 10$. Below this, the dependency is much weaker, hence why $\epsilon_{\rm z1,i1}$ is so close to $\epsilon_{\rm ideal}$. It is worth noting that for the very most nearby stars, even a quite large $z$ still has a quite small impact. The trail of data points at the bottom of the figure are due to different representations of the two most nearby stars ($\alpha$ Cen A/B). It can be seen that at the distance of 1.3 pc, the disk is so well-resolved that even at $z \sim$ 100, the disk brightness is still dominated by the planet brightness, and does not contribute significantly to the noise. The inclination dependency extends over the whole range of $i$. 

In this simulation, we have chosen a set of realistic parameters to provide an illustration of the effects of knowing the inclination and exozodiacal brightness. The exact numbers can obviously be expected to vary somewhat with changed mission parameters or underlying physical assumptions. Two important aspects -- the residual PSF level and the exozodiacal brightness distribution -- are discussed in the next section. Other mission parameters that may vary are the IWA or number of selected targets. The specific mission parameters also affect how $\bar{f}_{\rm vis}$ should be factored in. For instance, in the event of a coronagraphic mission and with a priori known phase angles (e.g. from SIM, see Sect. \ref{sec_large}), it is possible to schedule the observation such that it occurs when the planet is known to be outside of the IWA, and in this case, the present factoring of $\bar{f}_{\rm vis}$ is an over-correction. In the other extreme, if we have an occulter mission where the target-to-target motion dominates the fuel budget such that the target list and visit schedule must be strictly pre-defined, then it becomes relatively more important that there is a high chance that the target is visible than what the required integration time is going to be, and then the present factoring of $\bar{f}_{\rm vis}$ is an under-correction. Finally, specific target properties, such as binarity, exozodiacal dependence on mass etc. could also have some impact on the outcome. In any case, it is clear that the impacts of $i$ and $z$ are large, and that it is therefore highly useful to determine these quantities a priori, to the greatest extent that is technologically feasible.

\subsection{Alternative results}
\label{sec_alternative}

In the main simulation, we considered missions in which a contrast of $10^{-10}$ is reached outside of the IWA. However, since such a high contrast is technologically challenging, the idea is sometimes considered to loosen the constraints on the contrast of the residual stellar PSF, remove PSF substructure through differential imaging, and integrate against the residual random noise. Hence, we have also performed a calculation using a contrast of $10^{-9}$ for comparison. The simulation was otherwise run with identical parameters as the main simulation.

\begin{table}
   \centering
\caption[]{Alternative relative efficiencies in the event of high PSF residuals.}
         \label{inctab2}
\begin{tabular}{lll}
            \noalign{\smallskip}
            \hline
            \noalign{\smallskip}
$z$ knowledge & $i$ unknown & $i$ known \\
            \noalign{\smallskip}
            \hline
            \noalign{\smallskip}
$z > 1000$ & $\epsilon_{\rm hp,z3,i0} = 1.00$ & $\epsilon_{\rm hp,z3,i1} = 1.14$ \\
$z > 100$  & $\epsilon_{\rm hp,z2,i0} = 1.02$ & $\epsilon_{\rm hp,z2,i1} = 1.16$ \\
$z > 10$   & $\epsilon_{\rm hp,z1,i0} = 1.04$ & $\epsilon_{\rm hp,z1,i1} = 1.17$ \\
Ideal case & N/A                              & $\epsilon_{\rm hp,ideal} = 1.17$ \\
            \noalign{\smallskip}
            \hline
\end{tabular}
\end{table}

The results are shown in Table \ref{inctab2}, where the relative efficiencies are denoted with index `hp' to signify the high PSF case. Unsurprisingly, it can be seen that the impact of a priori knowledge decreases, though it remains significant, at least for the inclination. The decreased impact is due to the fact that the noise is to a much greater extent dominated by the residual stellar PSF, which depends on neither $i$ nor $z$. However, the total efficiency of the mission also goes down dramatically, since what this scenario does is raise the noise floor, and thus required integration time, for all targets. Due to the consequential rapid decrease of science-to-cost ratio with increasing residual PSF noise, it is dubious whether the high PSF case can ever be worthwhile for the purpose of a dedicated mission for finding Earth-like planets.

The most uncertain physical conditions in the simulations are the brightness distributions of exozodiacal disks, both in terms of the $z$ distribution and in terms of the brightness as function of $r$ in individual disks. For the $z$ distribution in the main simulations, we used a slope based on known 24 $\mu$m excesses, because it is a better representative of warm dust in the inner disk than 70 $\mu$m excesses. On the other hand, the number of Sun-like stars with known 70 $\mu$m excesses is larger, since it is an easier quantity to measure. The 70 $\mu$m excesses give a steeper slope of $\alpha = 1.08$ than the $\alpha = 0.80$ slope derived at 24 $\mu$m. For completeness, and to quantify the impact of this uncertainty, we run a set of simulations for the steeper slope case, denoted with index `st'. The resulting efficiencies are shown in Table \ref{inctab3}.

\begin{table}
   \centering
\caption[]{Alternative relative efficiencies in the event of a steep slope for the exozodiacal brightness distribution.}
         \label{inctab3}
\begin{tabular}{lll}
            \noalign{\smallskip}
            \hline
            \noalign{\smallskip}
$z$ knowledge & $i$ unknown & $i$ known \\
            \noalign{\smallskip}
            \hline
            \noalign{\smallskip}
$z > 1000$ & $\epsilon_{\rm st,z3,i0} = 1.00$ & $\epsilon_{\rm st,z3,i1} = 1.33$ \\
$z > 100$  & $\epsilon_{\rm st,z2,i0} = 1.23$ & $\epsilon_{\rm st,z2,i1} = 1.59$ \\
$z > 10$   & $\epsilon_{\rm st,z1,i0} = 1.46$ & $\epsilon_{\rm st,z1,i1} = 1.77$ \\
Ideal case & N/A                              & $\epsilon_{\rm st,ideal} = 1.81$ \\
            \noalign{\smallskip}
            \hline
\end{tabular}
\end{table}

In this case, the impact of prior knowledge goes up, this is mainly because there are simply a larger number of stars with high values of $z$ for the observer to identify and avoid. This also means that the relative importance of knowing $z$ compared to knowing $i$ goes up, for instance $\epsilon_{\rm st,z1,i0}$ is higher than $\epsilon_{\rm st,z3,i1}$. As mentioned above, we prefer the 24 $\mu$m slope for physical reasons, but on the other hand, it would be a disappointing experience if a large-scale dedicated mission for Earth-like planets was launched and it was found that the number of bright exozodiacal disks is larger than expected, so from a conservative viewpoint there are reasons to also consider the worst-case scenario. 

We also chose to use the Beichman et al. (2006) and Trilling et al. (2008) 24 $\mu$m excesses rather than those of Koerner et al. (2010), in order to avoid biasing the sample with the large number of K-type stars. On the other hand, the total Koerner et al. (2010) sample is larger, so in the event that it is representative, it is a better statistical indicator. Hence, we have also performed simulations with a slope of $\alpha = 0.36$, which is the best fit to the Koerner et al. (2010) distribution. The results are shown in Table \ref{inctab4}.

\begin{table}
   \centering
\caption[]{Alternative relative efficiencies in the event of a shallow slope for the exozodiacal brightness distribution.}
         \label{inctab4}
\begin{tabular}{lll}
            \noalign{\smallskip}
            \hline
            \noalign{\smallskip}
$z$ knowledge & $i$ unknown & $i$ known \\
            \noalign{\smallskip}
            \hline
            \noalign{\smallskip}
$z > 1000$ & $\epsilon_{\rm sh,z3,i0} = 1.00$ & $\epsilon_{\rm sh,z3,i1} = 1.15$ \\
$z > 100$  & $\epsilon_{\rm sh,z2,i0} = 1.00$ & $\epsilon_{\rm sh,z2,i1} = 1.15$ \\
$z > 10$   & $\epsilon_{\rm sh,z1,i0} = 1.01$ & $\epsilon_{\rm sh,z1,i1} = 1.15$ \\
Ideal case & N/A                              & $\epsilon_{\rm sh,ideal} = 1.16$ \\
            \noalign{\smallskip}
            \hline
\end{tabular}
\end{table}

As expected, the total benefit of increased knowledge decreases in this case and reaches values similar to those of the high PSF case. The inclination remains an important parameter still, but the benefit of knowing $z$ becomes practically negligible. Hence, it is clear that the uncertainty of the exact $z$ distribution leads to a broad possible range of mission impact of prior knowledge. From a risk minimization perspective, this in fact further underlines the importance of determining $z$ and $i$ a priori.

Regarding the distribution of the disk brightness as function of $r$, we chose a slope of $r^{-2.3}$, on the basis of visual light observations. On the other hand, more recent mapping at 24 $\mu$m is rather more indicative of a slope of order $r^{-4.5}$, implying that our knowledge of this distribution is incomplete. To take also this uncertainty into account, we perform yet another simulation with the $r^{-4.5}$ slope. The numbers are provided in Table \ref{inctab5}.

\begin{table}
   \centering
\caption[]{Alternative relative efficiencies in the case where the morphology of the zodiacal disk is based on the thermal brightness distribution.}
         \label{inctab5}
\begin{tabular}{lll}
            \noalign{\smallskip}
            \hline
            \noalign{\smallskip}
$z$ knowledge & $i$ unknown & $i$ known \\
            \noalign{\smallskip}
            \hline
            \noalign{\smallskip}
$z > 1000$ & $\epsilon_{\rm th,z3,i0} = 1.00$ & $\epsilon_{\rm th,z3,i1} = 1.28$ \\
$z > 100$  & $\epsilon_{\rm th,z2,i0} = 1.28$ & $\epsilon_{\rm th,z2,i1} = 1.59$ \\
$z > 10$   & $\epsilon_{\rm th,z1,i0} = 1.58$ & $\epsilon_{\rm th,z1,i1} = 1.86$ \\
Ideal case & N/A                              & $\epsilon_{\rm th,ideal} = 1.92$ \\
            \noalign{\smallskip}
            \hline
\end{tabular}
\end{table}

Since the distribution is much steeper in this case, the benefit of prior knowledge increases significantly, providing the highest numbers among the simulations performed here. Given the fact that the brightness distribution of our own zodiacal disk is not only poorly known, but in addition may not necessarily be representative of the typical exozodiacal disk in the first place, it is necessary to consider that the typical distribution could be as steep as considered here, and thus the impact of prior $i$ and $z$ knowledge could be as large.

\section{Inclination determinations}
\label{sec_inclin}

The fact that the inclinations of stellar systems are poorly known certainly indicates that such a determination is non-trivial. Beatty \& Seager (2010) discuss some different methods of inclination determination, here we will briefly discuss those for the context of a direct imaging survey as examined here, as well as discuss some additional methods. We separate the discussion between small-number methods, which can be used to determine the inclination in special cases, and large-number methods, which could potentially determine inclinations for the whole sample. The methods typically assume that the stellar rotation and planetary orbit are coplanar.

\subsection{Small-number methods}
\label{sec_small}

One method mentioned in Beatty \& Seager (2010) is $v \sin i$ and $P_{\rm rot}$ determinations to constrain $\sin i$. This is useful for particularly rapidly rotating stars, but probably not for typical stars due to an ambiguity with macroturbulence. The analysis by e.g. Saar \& Osten (1997) shows that even though the two effects are in principle distinguishable to some extent, the resulting $\sin i$ estimation is often of limited utility. A different method that could work in individual cases is determining the orbital plane of other circumstellar material, when such is present. We have already mentioned, e.g., the case of $\epsilon$ Eri where a spatially resolved dust disk is present (Greaves et al. 2005). Outer giant planet companions could be used for the same purpose. Such companions will be increasingly detectable over the coming years, with instruments like GPI or SPHERE, and in a longer perspective, JWST and ground-based ELT:s.

In principle, one might similarly expect that the orbital planes of multiple stellar systems could correspond to their rotational planes, and therefore the orbital planes of their respective planets. However, the formation and dynamical evolution of such systems leave uncertainties regarding the accuracy of inferring a rotational inclination from stellar orbital inclinations. We have tested this accuracy on a set of known multiple systems. As an input sample, we use the Sixth Orbital Catalog of Visual Binary Stars\footnote{http://ad.usno.navy.mil/wds/orb6.html} to select all stellar systems with three or more components. Out of these, we select only those systems where inclination and ascending node measurements exist with formal errors given, for two or more of the orbits in the system. One of the remaining systems is Sgr A, which we remove as it is clearly not applicable to our study. What remains is 10 systems, of which three are quadruple (HIP 21402, Lane et al. 2007; HIP 28614, Muterspaugh et al. 2008; HIP 76563, Drummond et al. 1995) and the rest are triple systems (HIP 2552, Docobo et al. 2008; HIP 9500, Hartkopf et al. 2000; HIP 79607, Raghavan et al. 2009; HIP 84949 and HIP 107354, Muterspaugh et al. 2008; HIP 113726, Hartkopf et al. 2009; HIP 116726, Olevic \& Cvetkovic 2005). The idea is to test whether at least the orbits within the system are coplanar with each other, which is a necessary pre-condition for a rotational plane to be inferred from any given orbit orientation.

The differential orientation $\Delta \zeta$ is calculated on the basis of spherical trigonometry from the differential inclination $\Delta i$ and ascending node $\Delta \Omega$:

\begin{equation}
\Delta \zeta = \arccos(\cos(\Delta i) \cos(\Delta \Omega))
\end{equation}

For the quadruple systems (except HIP 76563 where only two of the orbits are sufficiently constrained), two differential orientations are calculated, where the most precisely determined orbit is related to each of the other two orbits. The results are shown in Table \ref{inctab6}. It is immediately obvious that there is no strong preference towards coplanarity, although one system might be said to be conspicuously close to prograde coplanarity, and another one to retrograde coplanarity (but neither is perfectly coplanar within the errors). We test this more rigorously by generating $10^5$ orbital pairs with random orbits, following the expected distributions (composite of uniformly random ascending nodes and the expected distribution of inclinations, i.e. the arccosine of a uniform distribution), and calculate differential orientations as above. The randomly generated distribution is then compared to the observed distribution through a Kolmogorov-Smirnov test. The probability that the populations are drawn from the same distribution is 69\%. The population must therefore be described as being fully consistent with being random. Although there could hypothetically be some weak preference towards coplanarity, it is clearly far too small to make inferences about unseen inclinations in the systems.

\begin{table}
   \centering
\caption[]{Orientations of triple and quadruple systems with multiple known $i$ and $\Omega$.}
         \label{inctab6}
\begin{tabular}{ll}
            \noalign{\smallskip}
            \hline
            \noalign{\smallskip}
System name           & $\Delta \zeta$ \\
(Orbit 1 / Orbit 2)   & [deg]         \\
            \noalign{\smallskip}
            \hline
            \noalign{\smallskip}
HIP 2552 AaAb/AB      & 3.6$\pm$1.5   \\
HIP 9500 AB/AC        & 50.5$\pm$2.5  \\
HIP 21402 Aa1Aa2/AaAb & 125.9$\pm$2.0 \\
HIP 21402 Ab1Ab2/AaAb & 106.3$\pm$5.8 \\
HIP 28614 AaAb/AB     & 126.1$\pm$7.6 \\
HIP 28614 BaBb/AB     & 93.5$\pm$3.8  \\
HIP 76563 AB/CD       & 66.1$\pm$3.0  \\
HIP 79607 AB/BaBb     & 175.0$\pm$3.2 \\
HIP 84949 AB/BaBb     & 26.8$\pm$1.7  \\
HIP 107354 AB/BaBb    & 47.8$\pm$2.5  \\
HIP 113726 AB/AaAb    & 102.6$\pm$2.3 \\
HIP 116726 AB/BaBb    & 28.7$\pm$9.5  \\
            \noalign{\smallskip}
            \hline
\end{tabular}
\end{table}

Since the dynamical history of multiple systems might be more violent than for an ordinary binary, it cannot be categorically excluded that a binary orbital inclination could be a better probe for the rotational plane of each single stellar component. However, it is clear that the viability of such a method would have to be carefully proven before it could be applied in practice.

\subsection{Large-number methods}
\label{sec_large}

The use of asteroseismology to determine inclinations for a large number of targets in the context of transit probabilities is discussed in Beatty \& Seager (2010), particularly with regards to SONG (Grundahl et al. 2008), with an estimate that inclination determinations would require a time investment of approximately 10 nights per star. SONG is a planned network of 1-m class telescopes spread over all longitudes, equipped with high-resolution spectrographs for asteroseismology studies. For the transit survey envisioned in Beatty \& Seager (2010) with 830 stars, the total requirement becomes $\sim$23 years, which seems excessive. However, in the context of the imaging survey studied here involving 164 stars, we note that the requirement comes down to 4.5 years, which would be well within the timescale for a launch of a dedicated Earth-like planet search mission, and hence feasible. Therefore, if asteroseismology can indeed provide inclinations to a reasonable precision, it appears a network like SONG could provide the required precursor science.

A different possibility to determining inclinations for the bulk of the potential targets comes with the Space Interferometer (SIM; e.g. Unwin et al. 2008). SIM would not only provide the inclinations and even phase angles, but also identify which systems actually have planets in the first place, which is particularly useful if the frequency of Earth-like planets is low. On this note, it should be pointed out that SIM is in fact blind to the types of planets we have assumed for our simulations, since they would have a 1-year period, which is fundamentally indistinguishable from parallax errors. In any case, SIM would be hugely useful in determining inclinations for all stars that have any sufficiently massive planets. Taking the argument one step further, however, it might even be argued that since a main point of SIM would be to detect planets that are suitable for further study with direct imaging and spectroscopy, the target priorities of SIM should by themselves be partly influenced by a priori determinations of the inclinations, since those systems that are closest to face-on will be those that are easiest to study with imaging techniques.

\subsection{Additional utility of inclination}
\label{sec_utility}

In this paper, we have mainly focused on one strict utility from knowing planetary system inclinations a priori -- maximizing the efficiency outcome from selection of targets for the detection of extrasolar Earth-like planets. However, knowing the inclination (and selecting for low-inclination systems) is useful for even wider purposes related to the study of such planets. With an a priori known inclination, it becomes easier to formulate a re-visit strategy once a given planet is detected, and for a close to face-on system, that schedule is minimally strict. Photometric or spectroscopic study of long-term atmospheric trends, in particular seasonal variations, benefits from low inclinations to maximize the continuous baseline. It might be argued that day-night variations would be harder to study in low-inclination systems, if the obliquity is close to zero. However, Earth's obliquity is significantly non-zero, as are the Solar system planets in general, so this is unlikely to be a limiting factor. In addition, interaction of an Earth-like planet with an exozodiacal disk is likely to cause disk structures, which depend on the dynamical mass of the planet, thus potentially providing a mass estimate independently of astrometry or radial velocity. Such structure is most easy to characterize at low inclinations, since ambiguity along the line of sight starts to occur at high inclinations. Finally, knowing the inclination has positive spin-off effects on other areas of exoplanet studies such as for determining transit probabilities as in Beatty \& Seager (2010), or for yielding real mass estimates for radial velocity planets.

\section{Conclusions}
\label{sec_conclusions}

We have studied the impact of knowing the inclinations and/or exozodiacal brightnesses for nearby stars on the efficiency of detection of extrasolar Earth-like planets. It was found that for a 60 mas IWA and a selection of 60 out of the 164 most nearby Sun-like stars, the efficiency can be boosted by 23\% by knowing the inclinations of all the systems, even if only $>$1000 zodi disks are known (as is the present status). If all $>$10 zodi disks are known but no inclinations, the boost is 12\%. A combination of the two provides a 32\% improvement. These improvements may be smaller or larger depending on the exact distribution of exozodiacal brightnesses, and some mission parameters. Simultaneously determining all $>$10 zodi disks and all inclinations gives a near-ideal target selection. Attaining such a state of knowledge requires substantial effort but is probably realistic, and should be seriously considered as precursor science for dedicated direct imaging missions. 

Although $v \sin i$ measurements and circumstellar material can provide inclination constraints for a fraction of the systems, it is probably necessary to use asteroseismology or space-based astrometry to provide inclination measurements for the bulk of nearby Sun-like stars. The orbital planes in multiple stellar systems is however not an acceptable proxy for the rotational planes of individual stars or the orbital planes of their planets.

\section*{Acknowledgments}

M.J. is supported through the Reinhardt postdoctoral fellowship from the University of Toronto. The study made use of the CDS and NASA/ADS databases.

\label{lastpage}


\begin{thebibliography}{99}
\bibitem[2008]{Absil08} Absil, O., Akeson, R., Ardila, D. et al. 2009, ``2008 Exoplanet Forum Report'', Eds. P.R. Lawson, W.A. Traub, \& S. C. Unwin, http://exep.jpl.nasa.gov/documents/Forum2008\_109\_small.pdf
\bibitem[2007]{Agol07} Agol, E. 2007, MNRAS 374, 1271
\bibitem[2010]{Beatty10} Beatty, T.G. \& Seager, S. 2010, ApJ 712, 1433
\bibitem[2005]{Beck05} Beck, J.G. \& Giles, P. 2005, ApJ 621, L153
\bibitem[2003]{Beichman03} Beichman, C.A. 2003, ``Towards Other Earths: DARWIN/TPF and the Search for Extrasolar Terrestrial Planets'', Eds. M. Fridlund, Th. Henning, ESA SP-539, 271
\bibitem[2006]{Beichman06} Beichman, C.A., Bryden, G., Stapelfeldt, K.A. et al. 2006, ApJ 652, 1674
\bibitem[2006]{Cash06} Cash, W. 2006, Nature 422, 51
\bibitem[2009]{Cockell09} Cockell, C.S., L\'eger A., Fridlund M. et al. 2009, Astrobiology 9, 1
\bibitem[1994]{Dermott94} Dermott, S.F., Jayaraman, S., Xu, Y. et al. 1994, Nature 369, 719
\bibitem[2008]{Docobo08} Docobo, J.A., Tamazian, V.S., Balega, Y.Y. et al. 2008, A\&A 478, 187
\bibitem[1995]{Drummond95} Drummond, J.D., Christou, J.C., \& Fugate, R.Q. 1995, AJ 450, 380
\bibitem[2009]{Fabrycky09} Fabrycky, D.C. \& Winn, J.N. 2009, ApJ 696, 1230
\bibitem[2005]{Greaves05} Greaves, J.S., Holland, W.S., Wyatt, M.C. et al. 2005, ApJ 619, 187
\bibitem[2010]{Greaves10} Greaves, J.S. \& Wyatt, M.C. 2010, MNRAS accepted
\bibitem[2008]{Grundahl08} Grundahl, F., Christensen-Dalsgaard, J., Arentoft, T. et al. 2008, Communications in Asteroseismology 157, 273
\bibitem[2006]{Guyon06} Guyon, O., Pluzhnik, E.A., Kuchner, M.J., Collins, B, \& Ridgway, S.T. 2006, ApJS 167, 81
\bibitem[1976]{Hanner76} Hanner, M.S., Weinberg, J.L., Beeson, D.E., \& Sparrow, J.G. 1976, ``Interplanetary dust and zodiacal light'', Proc. of the colloqium, 31st, Heidelberg, West Germany, p.29-35
\bibitem[2000]{Hartkopf00} Hartkopf, W.I., Mason, B.D., McAlister, H.A. et al. 2000, AJ 119, 3084
\bibitem[2009]{Hartkopf09} Hartkopf, W.I. \& Mason, B.D. 2009, AJ 138, 813
\bibitem[2007]{Janson07} Janson, M. 2007, PASP 119, 214
\bibitem[2010]{Koerner10} Koerner, D.W., Kim, S., Trilling, D.E. et al. 2010, ApJ 710, L26
\bibitem[2007]{Lane07} Lane, B.F., Muterspaugh, M.W., Fekel, F.C. et al. 2007, ApJ 669, 1209
\bibitem[2009]{LeBouquin09} Le Bouquin, J.-B., Absil, O., Bensity, M. et al. 2009, A\&A 498, L41
\bibitem[1981]{Leinert81} Leinert, C., Richter, I., Pitz, E. \& Planck, B. 1981, A\&A 103, 177
\bibitem[2008]{Muterspaugh08} Muterspaugh, M.W., Lane, B.F., Fekel, F.C. et al. 2008, AJ 135, 766
\bibitem[2009]{Narita09} Narita, N., Sato, B., Hirano, T., \& Tamura, M. 2009, PASJ 61, L35
\bibitem[2010]{Nesvorny10} Nesvorn\'y, D., Jeniskens, P., Levison, H. et al. 2010, ApJ 713, 816
\bibitem[2005]{Olevic05} Olevic, D. \& Cvetkovic, Z. 2005, Serbian AJ 170, 65
\bibitem[1997]{Perryman97} Perryman, M.A.C., Lindegren, L., Kovalevsky, J. et al. 1997, A\&A 323, 49
\bibitem[2009]{Raghavan09} Raghavan, D., McAlister, H.A., Torres, G. et al. 2009, ApJ 690, 394
\bibitem[1997]{Saar97} Saar, S.H. \& Osten, R.A. 1997, MNRAS 284, 803
\bibitem[2008]{Trilling08} Trilling, D.E., Bryden, G., Beichman, C.A. et al. 2008, ApJ 674, 1086
\bibitem[2008]{Unwin08} Unwin, S.C., Shao, M., Tanner, A.M. et al. 2008, PASP 120, 38
\bibitem[2009]{Winn09} Winn, J., Johnson, J.A., Fabrycky, D. et al. 2009, ApJ 700, 302
\end{thebibliography}
\end{document}